\documentclass[aps,preprint,showpacs,amsmath,amssymb,floatfix,pre]{revtex4}
\usepackage{graphicx}
\usepackage{dcolumn}
\usepackage{bm}

\begin{document}

\title{Understanding the Unusual Properties of Water}

\author{Giancarlo FRANZESE$^1$ and H.~Eugene STANLEY$^2$}
\address{
$^1$Departament de F\'{\i}sica Fonamental, Universitat de Barcelona\\
Diagonal 647, 
08028 Barcelona, Spain,
E-mail: gfranzese@ub.edu\\
$^2$Center for Polymer Studies and Department of Physics, \\ 
Boston University,
Boston, MA 02215 USA
E-mail: hes@bu.edu}  

\date{fs-21-02-06.tex ~~~~ 21 February 2006}

\begin{abstract}
Water is commonly associated to the existence of life.  However, there
is no clear reason why water should be the only liquid in which life
could form and survive.  Since the seminal work of L. J. Henderson in
1913, scientists are trying to answer the question about the relation
between the unusual properties of water and the existence of life.
Here we follow the first steps along the challenging path to this
answer, trying to understand (i) what is unusual about water; (ii) why
water has anomalies; (iii) which are the full implications of these
unusual properties; and (iv) if these anomalies are exclusive
properties of water.  By identifying some interesting clues, then by
formulating a working hypothesis and, next, by testing it, we find the
surprising results that some properties of water, such as the
anomalous behavior of density, are very sensitive to small changes of
the microscopic interactions, while others, such as the presence of
more than one crystal or a possible phase transition between two
liquids with different local structures, are predicted not only in
water, but also in other liquids.  Since the change of local liquid
structure could be relevant in biological processes, the possibility
of a wide class of liquids with this property could help in
understanding if water is essential for life.
\end{abstract}

\maketitle

\section{Introduction}

Water is commonly associated to the existence of life.  
It is the main component of living organisms: the human body is by
weight roughly 75\% water in the first days of life and 
roughly 60\% water in the adult age. The majority of this water
(roughly 60\%) is inside the cells, while the rest (extracellular
water) flows in the blood and below the tissues.  

Many living beings can survive without water only a few days.
This is because water participates in the majority
of the biological processes \cite{Franks1985}, such as the
metabolism of nutrients catalysed by enzymes. In order to be
effective the enzymes need to be suspended in a
fluid to adopt their 
active three-dimensional structure. The main reason why 
we need water is that
it allows the processes of elimination of cellular metabolic
residues. Is through the water that our cells can communicate and 
that oxygen and nutrients can be brought to our tissues. 

However, despites all these considerations, there is no clear reason why
water should be the only liquid in which life could form and survive.
The debate about the essential role in biological processes 
played by water and its properties is open.

The first remarkable thing about water is that it has
a large number of {\it anomalies} with respect to the other
usual substances. This fact has fascinated the scientists since centuries.
It was, probably, L. J. Henderson in 1913
\cite{Henderson} the first asking about the relation between the 
unusual properties of water and the existence of life.

Since then, scientists are trying to reach the answer to this question step by
step. Here we will follow the first
steps of this long path. Those steps can be summarized by the following
questions: 
\begin{itemize}

\item
What is unusual about water? 

\item
Why water has anomalies?

\item
Which are the full implications of these unusual properties?

\item
Are these anomalies exclusive properties of water?

\end{itemize}

Answer to these questions is a scientific
challenge. We will organize the discussion by, first, identifying some
interesting clues, then by formulating a working hypothesis and, next, by
testing it. In this process we will use, among the other theoretical
tools, computer simulations. Their applicability to the case of water
has been demonstrated by many studies and the dramatic increase of
computational power makes them more and more feasible.

\section{What is unusual about water?}

There are many puzzles associated with water. Their relevance for
biological processes is sometimes difficult to say, but there is one
that is definitely evident: liquid water can exist at temperatures far
lower than the melting/freezing temperature.

This property allows living
systems to survive at very cold temperatures, far below 0$^\circ$C. At this
temperatures liquid water is said to be in a {\it supercooled} phase,
to emphasize that this phase is metastable, i. e. it cannot last
forever as a stable phase, but, sooner or later, it will transform
into ice.  

\subsection{Water at very low temperatures}

Water in
organic cells can avoid freezing at temperatures as low as
-20$^\circ$C in insects, and -47$^\circ$C in plants, for a time long
enough to be compared to their lifetime \cite{debenedettibook}.  In
laboratory supercooled water has been observed at -41$^\circ$C at
ambient pressure \cite{Cwilong1947} and {\it in situ} observation has
confirmed the presence in clouds of droplets of liquid water at
-37$^\circ$C \cite{Rosenfeld00}. The lowest measured supercooled
liquid water temperature is -92$^\circ$C at 2 kbars \cite{Kanno1975}.

Below these temperatures water cannot 
exist as a liquid and forms crystalline ice, by means of a process
known as {\it crystal homogeneous nucleation}. But water is unusual also
in its crystalline phase. Indeed, water is a {\it polymorph},
i. e. has many crystalline forms, like a few other unusual substances,
e. g. carbon, which may form the cheap graphite or the very expensive
diamond. In the case of water, the 
known polymorphs are as much as thirteen.
 
At roughly 80 degrees below the homogeneous nucleation
temperature water can exists also in a glassy form, i. e. in a frozen,
amorphous solid form \cite{BrugellerMayer1980}, which crystallize to
cubic ice at approximately -123$^\circ$C at ambient pressure. 
Again, unusually, instead of just one
amorphous phase, at these temperatures there is more than one {\it
polyamorph} of glassy water.  The first clear indication of this was a
discovery of Mishima in 1984. At low pressure there is one form,
called low-density amorphous (LDA) ice \cite{BrugellerMayer1980},
while at high pressure there is a new form called high-density
amorphous (HDA) ice \cite{Mishima1985}, with a volume discontinuity
between these two phases of 27\%, comparable to that separating
low-density and high-density polymorphs of crystalline ice
\cite{Mishima1994-1996,Mishima96, SuzukiMishima2002,ms98}. 

Recently, a new form of amorphous water has been experimentally
observed by compressing HDA over 0.95~GPa, the very-high-density
amorphous (VHDA) ice \cite{Loerting01,Finney02}, with a volume
discontinuity between HDA and VHDA of 11\%.  
At 125~K (-148$^\circ$C) the three polyamorphs
can be formed by compression in a stepwise process LDA-HDA-VHDA
\cite{Loerting06}. 
However, from experimental data is not yet clear 
if the transformations LDA-HDA and HDA-VHDA are real discontinuous
transitions or just very sharp increases of densities and
additional
investigations are needed to clarify this point.
Remarkably,
HDA ice is the most abundant ice in the
universe, where it is found as a frost on interstellar grains
\cite{Jenniskens}. HDA ice has been proposed as the cradle where small
inorganic molecules combine into the large organic molecules 
at the origin of life \cite{webNasa}.

\subsection{Volume fluctuations}

The unusual behavior of water is not limited to the supercooled
phase. For example, its isothermal compressibility $K_T$, that is the
departure $\delta \bar{V}$ of the volume per particle $\bar{V}$ from
its mean value in response to an infinitesimal pressure change $\delta
P$, is anomalous already at high temperatures.

\begin{figure}
\includegraphics[width=5cm]{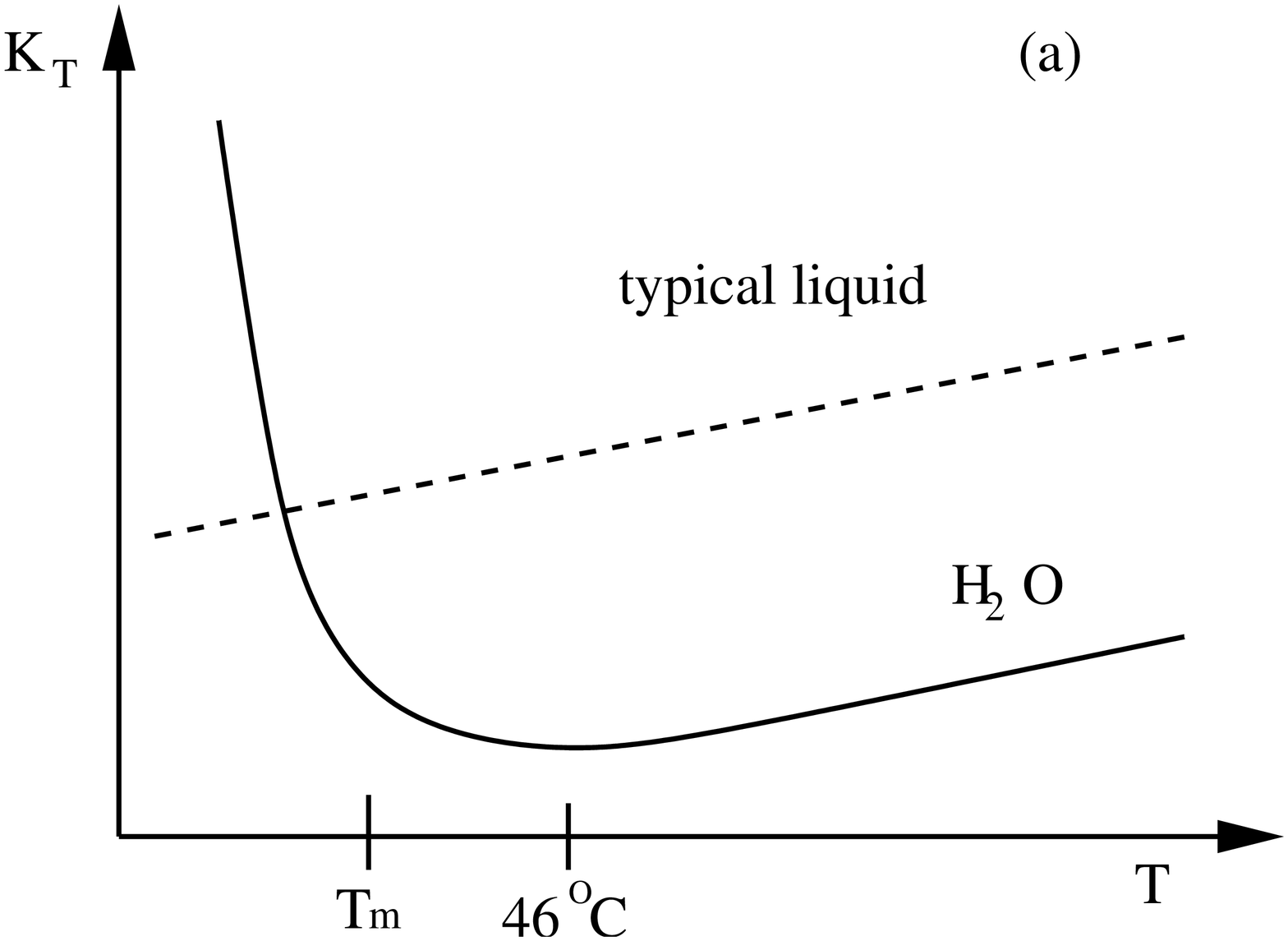}
\includegraphics[width=5cm]{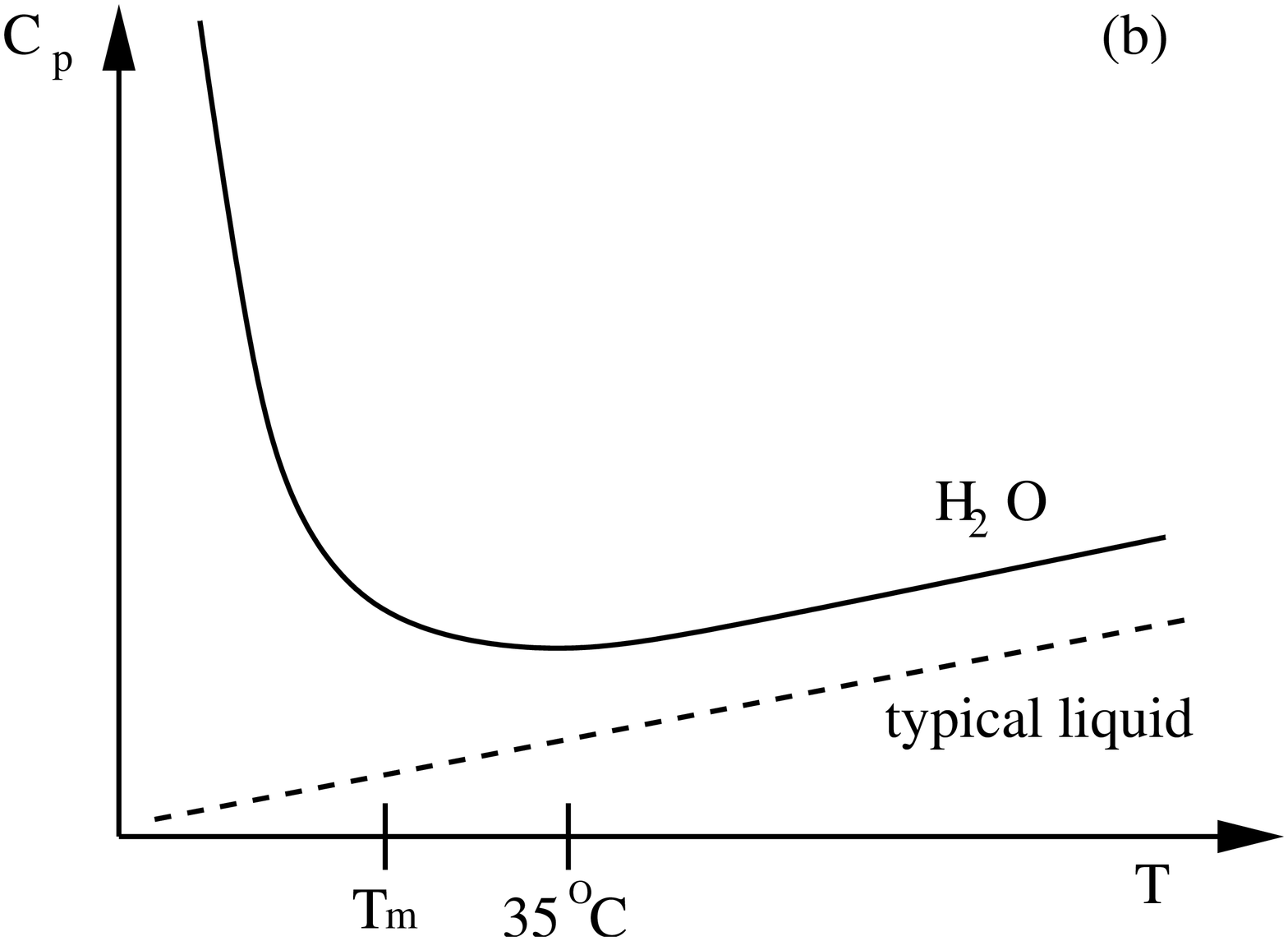}
\includegraphics[width=5cm]{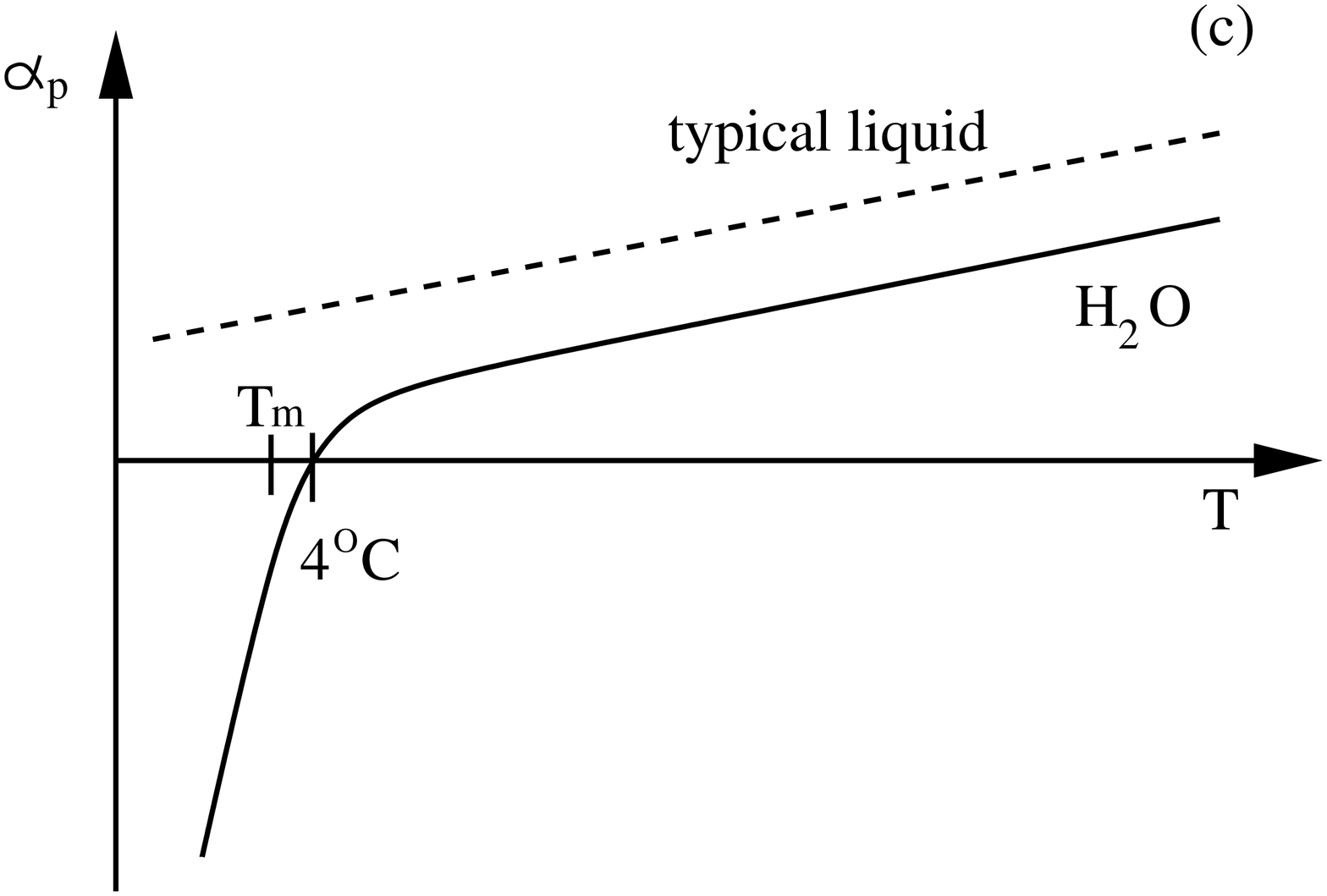}
\caption{Schematic dependence on temperature, at ambient pressure,  of
(a) the isothermal 
compressibility $K_T$ proportional to volume fluctuations, 
(b) the constant-pressure specific heat $C_P$ proportional to entropy
fluctuations,  
and (c) the thermal expansivity $\alpha_P$ proportional to volume-entropy
cross fluctuations $\langle \delta V \delta S \rangle$.
The behavior of a typical
liquid is indicated by the dashed line, which, very roughly, is an
extrapolation of the high-temperature behavior of liquid water. Note
that while the anomalies displayed by liquid water are apparent above
the melting temperature $T_m$, they become more striking as one
supercools below $T_m$.
\label{schem}}
\end{figure}

For a typical liquid $K_T$ decreases when one lowers the temperature
(Figure \ref{schem}).  In statistical physics we learn that $K_T$ is
proportional to the average value of the fluctuations of 
$\bar{V}$, hence we expect that $K_T$ decreases with the temperature $T$,
because the fluctuations decrease with $T$.  For water, instead, $K_T$
is anomalous in three respects: (i) it is larger than one would
expect; (ii) below $46^\circ$C, instead of decreasing as for usual
liquids, it increases, doubling its value before reaching the
homogeneous crystal nucleation temperature in the supercooled phase;
(iii) it appears as if it might diverge to infinity, at a temperature
of about $-45^\circ$C, increasing like a power law and hinting at some
sort of critical behavior \cite{Speedy76}.

\subsection{Entropy fluctuations}

Another anomalous thermodynamic function is the heat capacity at constant
pressure $C_P$, which is the response $\delta \bar{S}$ of the entropy
(or disorder) per particle to an
infinitesimal temperature change $\delta T$. This quantity decreases
with $T$ for a typical liquid, because it is proportional to the
fluctuations of entropy and these fluctuations decreases with $T$ for
normal substances. But water is, again, anomalous in the same three
respects: (i) its $C_P$ is larger than expected; (ii) below about 35~$^\circ$C
the specific heat increases; (iii) this increase is approximated by a
power law \cite{Angell82}. 
It is thanks to the high heat capacity of water that we can easily regulate
our body temperature by transpiration.

\subsection{Volume-entropy cross fluctuations}

The last anomalous thermodynamic function we consider is the thermal
expansivity $\alpha_P$, which is the response $\delta \bar{V}$ of the volume
to an infinitesimal temperature change $\delta T$. In typical liquids
this quantity decreases with $T$ and is always positive, because it is
proportional to the cross-fluctuations of entropy and volume $\langle
\delta \bar{V} \delta \bar{S}\rangle$. Indeed, in normal liquids 
this quantity is positive because when there is a local
increase of fluctuation of the volume $\delta \bar{V}$,
the particles in that region shuffle around increasing the disorder and the  
associated entropy fluctuation $\delta \bar{S}$. 
Water, again, is anomalous in the same three respects: (i)
its $\alpha_P$ is 3 time smaller than expected; (ii) below 4~$^\circ$C
its $\alpha_P$ is negative and grows rapidly in absolute value; (iii)
this absolute value increases as a power law \cite{Hare86}.

The fact that $\alpha_P$ is negative below 4$^\circ$C shows that water
is more disordered when it is more dense. For this reason ice melts if
its density increases, e. g. for an increase of pressure, because only
by disordering, i. e. liquefying, it can reach the desired
density. This behavior is related to the most famous anomaly of water:
its density maximum at 4~$^\circ$C. Below this temperature the volume of
water expands explaining why frozen water pipes break or 
why ice cubes float on water once liquid water cools down. In contrast,
solid forms of typical substances are denser than their liquid
form. Is thank to this anomaly that lakes and seas start to freeze
from top, allowing fishes to survive in the liquid water below the
ice.

\section{Why water has anomalies?}

After this brief introduction on the unusual properties of water, any
scientist would ask why water is so special. Perhaps, the first clue
dates back to Linus Pauling \cite{Pauling39}, who recognized
that the distinguish feature of water, compared to other chemically
similar substances, is the preponderance of hydrogen bonds.

Each water molecule has two hydrogen atoms and two lone electron pairs. 
Each hydrogen has a partial positive
charge, forming an O--H bond with en electron pairs on the
oxygen side.
Each lone pair tends to form a hydrogen bond with the
hydrogen of a nearby H$_2$O molecule, whose O--H bond points to the lone
pair. Hence, in a simplified view, each molecule can form four hydrogen bonds
attracting four nearby molecules.

Since the four electron pairs (two lone pairs and two pairs of the
O--H bond) repel each other, the four hydrogen bonds point 
along the vertexes of an almost perfect tetrahedron. In liquid water
many of the possible hydrogen bonds between nearby molecules are formed, giving
rise to a hydrogen bond tetrahedral network. The network is not static since
hydrogen bonds have a very short lifetime, of the order of
picoseconds, allowing 
the rotation and the 
diffusion of H$_2$O molecules \cite{Laage-Hynes2006}. This dynamic hydrogen bond network
slows down when the temperature is
decreased and freezes in a full hydrogen bonded network, below the homogeneous
nucleation temperature, transforming into ice. The reason why water
expands when forms ice at ambient pressure is because the
distance between nearby molecules in the ice tetrahedral network is
larger than the average inter-molecular distance in the liquid, giving
rise to an {\it open} structure (hexagonal ice or ice I$_h$).

Experiments \cite{Soper-Ricci00} show that this open tetrahedral structure is
preserved even in the liquid at $T$ as low as 268~K (-5$^\circ$C) up
to the second shell of molecules, and that this open structure can
coexist with a more compact structure in which the second shell
collapses, locally increasing the density of the liquid. This two kinds
of local structures are called low-density liquid (LDL), for the open
structure, and high-density liquid (HDL) for the collapsed structure,
in analogy with the LDA and HDA ices.  They were first observed in
computer simulations \cite{Poole} and have suggested two ways to develop a
coherent picture of the unusual behavior of water: (i) the {\it
liquid-liquid phase transition} hypothesis~\cite{Poole} and (ii) the
{\it singularity-free} interpretation~\cite{StanleyTeixeira80,Sastry96}.

\subsection{The liquid-liquid phase transition hypothesis}

\begin{figure}
\includegraphics[width=13cm]{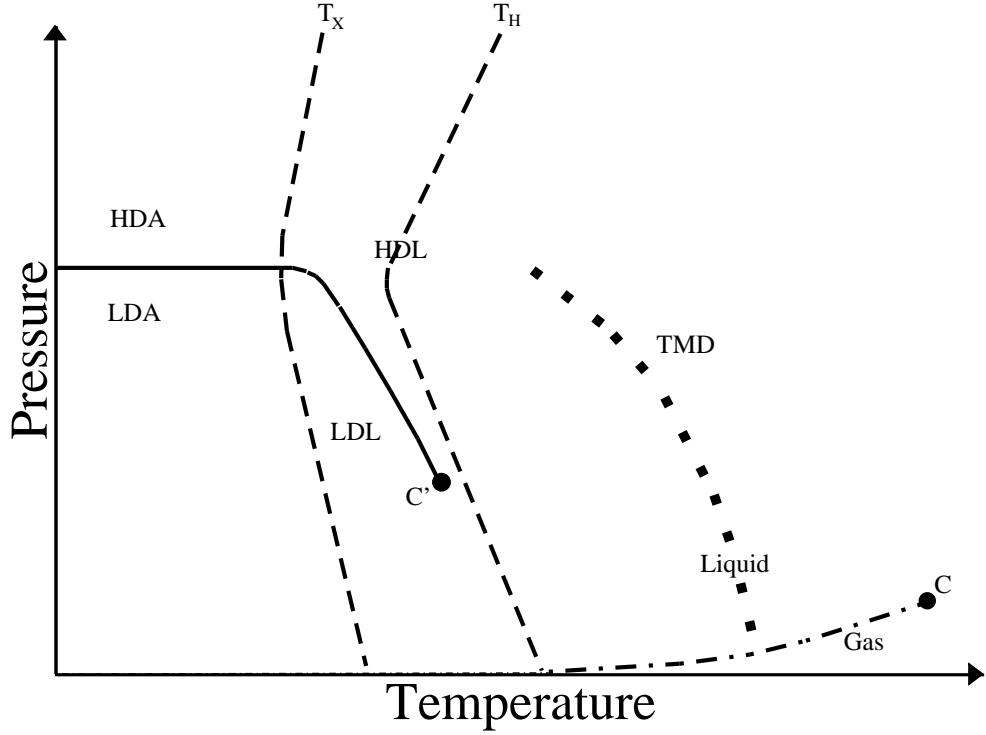}
\caption{Schematic representation of the 
liquid-liquid phase transition hypothesis
in the pressure $P$, temperature $T$ plane.
Below the temperature of spontaneous crystallization $T_X$, HDA ice
and LDA ice are separated by a first order phase transition
(continuous line). When this line reaches $T_X$ the slope of the
crystallization line changes and HDA and LDA transform continuously in
HDL and LDL, respectively. The coexistence line between the two
liquids ends in a critical point $C'$ below the homogeneous
crystallization temperature $T_H$. 
The dotted line denotes the temperature of
maximum density (TMD) in the liquid phase. The point-dotted line
represents the gas-liquid coexistence line ending in the critical
point $C$.
\label{schemPT}}
\end{figure}

In the liquid-liquid phase transition picture~\cite{Poole} the
LDA--HDA transition in supposed discontinuous and marked
by a line in the pressure--temperature ($P$--$T$) phase diagram
below the temperature $T_H$ of crystallization to cubic ice 
(Figure \ref{schemPT}). It is assumed that this line does
not terminate when it reaches the region of spontaneous
crystallization at $T_X$,
but extends into it, with LDA transforming without discontinuity into
LDL, and HDA into HDL, giving a LDL--HDL phase transition
line. According to this hypothesis, along this line the 
two kinds of liquids coexist as two separate phases, both
forming macroscopic droplets of a phase into the other. At
pressures above the liquid-liquid coexistence line the only
existing phase is HDL, while below there is only LDL. 
The collapsed structure of HDL is more disordered than the open
structure of LDL, hence the entropy of HDL is larger than that of LDL
and this implies, for thermodynamics relations, that the liquid-liquid
coexistence line has negative slope in the $P$--$T$ phase diagram.

An experiment able to cross the liquid-liquid coexistence line
should measure a sudden discontinuity in the local density of the liquid.
Theoretical and computational estimates of the liquid-liquid coexistence line
locate it in the region below the homogeneous nucleation of the crystal
at $T_H$ 
and above the spontaneous crystallization line at $T_X$. In this range
of temperatures the experiments
on the liquid cannot be performed, but the phase diagram can be
explored with the help
of computers and sophisticated water-models. These
simulations~\cite{Poole,geiger} show the discontinuity in density
which supports the hypothesis. They also show that the density difference,
between the two phases, decreases by increasing the temperature along
the line and goes to zero in a liquid-liquid {\it critical} point $C'$ 
where the line ends. At higher $T$ the two
phases are indistinguishable and the two kinds of structures (open and
collapsed) are found only at the microscopic scale of a few molecules,
just as the gas and liquid phases are indistinguishable above the
gas-liquid critical point $C$.

The existence of a critical point induces large ({\it critical})
fluctuations in a region that extends to temperatures and pressures
far away in the phase diagram. For example, experiments show that the
effect of the gas-liquid critical point $C$ on the response functions
is evident even at temperatures twice higher than the $C$-temperature
and that these functions diverge as power laws at $C$.
 
In a similar way, the anomalous increase of the response functions is,
in this hypothesis, the effect of approaching the liquid-liquid coexistence line,
with a genuine divergence at the critical point $C'$.  At $T$ above
the liquid-liquid critical temperature the thermodynamic response functions
appear to diverge to infinity approaching the Widom line, that is
defined as the analytic extension of the liquid-liquid phase transition line in
the $P$--$T$ phase diagram.  When the system is {\it extremely} close
to the Widom line, the functions will round off and ultimately remain
finite---as experimentally observed in the adiabatic compressibility
\cite{Trinh}.

This hypothesis is consistent with recent experiments
\cite{ms98}. The melting line of metastable ice IV and
stable ice V show an abrupt change in 
their slope as predicted if it would intersect the metastable liquid-liquid phase
transition line. However, the experiment resolution does not allow to
conclude if the sharp change is a real discontinuity, as required by 
the liquid-liquid phase transition hypothesis. Therefore, Mishima and one of us
(Stanley) interpolated the experimental data to calculate the Gibbs
free energy of the liquid at equilibrium with the different ice
polymorphs along their melting line and estimated a liquid-liquid critical point
at 1~kbar and 220~K.

\subsection{The singularity-free interpretation}

\begin{figure}
\includegraphics[width=13cm]{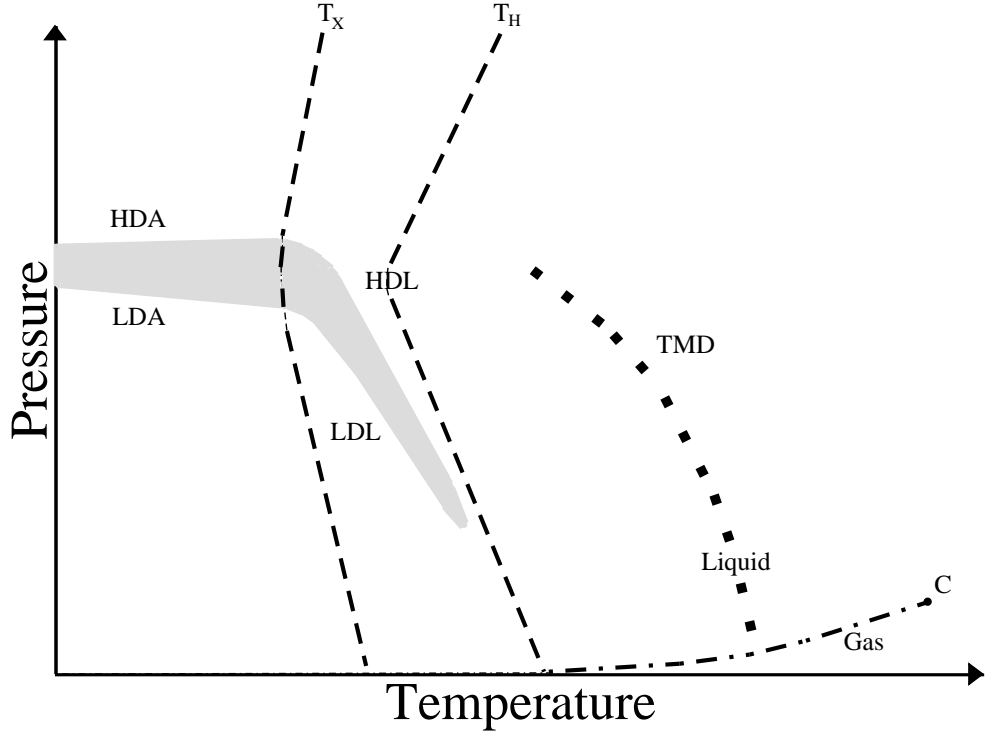}
\caption{Schematic representation of the singularity-free interpretation
in the pressure $P$, temperature $T$ plane.
The gray area represent the region where occurs a sharp, but continuous,
variation of density between HDA ice
and LDA ice, below $T_X$, and between HDL and LDL above $T_X$ and below
$T_H$. 
The dotted line denotes the temperature of
maximum density (TMD) in the liquid phase. The point-dotted line
represents the gas-liquid coexistence line ending in the critical
point $C$.
\label{schemPT2}}
\end{figure}

In the singularity-free scenario~\cite{StanleyTeixeira80,Sastry96}
(Figure \ref{schemPT2}) the LDL and HDL are 
still smoothly connected to LDA and HDA, respectively, but do not
represent two distinct phases separated by a discontinuous
transition. Instead, they represents local fluctuations of densities
and, by increasing the pressure, one can pass from LDL to HDL
observing a sharp, but continuous, increase of density, occurring in a
limited region. Hence, there
is no liquid-liquid phase transition and no liquid-liquid critical point in this scenario.
The large increase of response functions seen in the 
experiments represents only an {\it apparent} singularity that
eventually rounds off in a maximum, instead of giving rise to a real
divergence, in correspondence to the continuous increase of density.

This can be understood because, when LDL regions forms at low $T$, the
volume per particle increases $\delta \bar{V}>0$ while the entropy
(disorder) per particle decreases $\delta \bar{S}$ (the molecules are
ordered in a more extended tetrahedral structure) giving negative
$\langle \delta \bar{V} \delta \bar{S}\rangle$.  As we have seen, this
cross-fluctuation is proportional to the thermal expansivity
$\alpha_P$, that, as a consequence becomes negative at low 
temperature. In the same way, the anticorrelated
fluctuations of volume and entropy for LDL regions reduce $\alpha_P$
at high $T$ with respect to the expected value for an usual liquid.
Therefore, the unusual behavior of 
$\alpha_P$ can be easily interpreted in this scenario.

The temperature where $\alpha_P=0$ corresponds to the temperature of
maximum density (TMD) and, by changing $P$,
it form in the $P$--$T$ phase diagram a continuous line that bends
toward low $T$ when $P$ increases, as shown by experiments. All the
other anomalies we have mentioned so far can be interpreted as a
consequence of the presence of the TMD line.
Models and calculations have been shown to be consistent with this
scenario under some hypothesis, such as neglecting the correlations in
the hydrogen-bond formation and breaking.

\subsection{The difference between the two scenarios}

The liquid-liquid phase transition hypothesis and the singularity-free
interpretation are both thermodynamically consistent and
only differ in the experimentally-unaccessible
region, between $T_X$ and $T_H$.
Away from this region these two interpretations coincide, giving the
same description of the TMD line and the unusual behavior of the
response functions. 

However, the presence of a new liquid phase at low $T$ would represent
a relevant feature to understand the water dynamics not only in the
supercooled and glassy region, but also at moderately low
temperature, and it could be relevant for biological process, such as
the cold denaturation of protein \cite{manuel}. For example, 
the enzymatic activity of most proteins ceases below 220~K
\cite{Rasmussen92,Ferrand93}, the same
temperature estimated for the liquid-liquid critical point of water
\cite{ms98}, showing a possible relation between the two
phenomena. This is in agreement with the idea that the dynamics of
interfacial water and protein are highly coupled at low temperatures
\cite{masaki,Bagchi2005}. Further investigation is needed to understand
this relation.

\section{Which are the full implications of the unusual properties of water?}

Since from the experiments is difficult to reach a conclusive picture
about the origin of the anomalies of water, we adopt a different
approach, quite common in physics: we develop a schematic model that
reproduce the water properties and explore, by theoretical calculations
and computer simulations, its phase diagram, focusing on the
supercooled region hard to investigate with experiments.
The answers we can get in this way are not definitive, because
they depend on the inevitable approximations included in the model; hence,
different models can give different answers \cite{debenedetti-review}.
However, by studying how the answers change when we modify the features of
the model, we can help in clarifying which are the implications of the 
properties of water.

\subsection{The correlation of the hydrogen bond network}

We consider a model for water that (i) has the density anomaly, (ii) 
forms a hydrogen bond network, and (iii) has a parameter that describes how
strong is the correlation among the hydrogen bonds formed by the same molecule
\cite{fs,fsPhysA,fms}. The latter feature is introduced because 
experiments show that the relative orientations of the hydrogen bonds of a water
molecule are correlated, with the average H-O-H angle equal to 
$104.45^\circ$ in an isolated molecule, $104.474^\circ$ in the gas and 
$106^\circ$ in the high-$T$ liquid 
\cite{Kern-Hasted-Ichikawa,Hasted72,Ichikawa91}. 
Therefore, there is an ({\it intra-molecular}) interaction between the
hydrogen bonds formed by the same 
molecule. This interaction depends on 
$T$, because the H-O-H angle changes with temperature, consistent with
ab-initio calculations \cite{Silvestrelli} and molecular dynamics
simulations \cite{raiteri,Netz}. The strength of this intra-molecular 
interaction
represents the parameter we use in the model to regulate the
correlation among the hydrogen bonds of the same molecule.

\begin{figure}
\includegraphics[width=12cm]{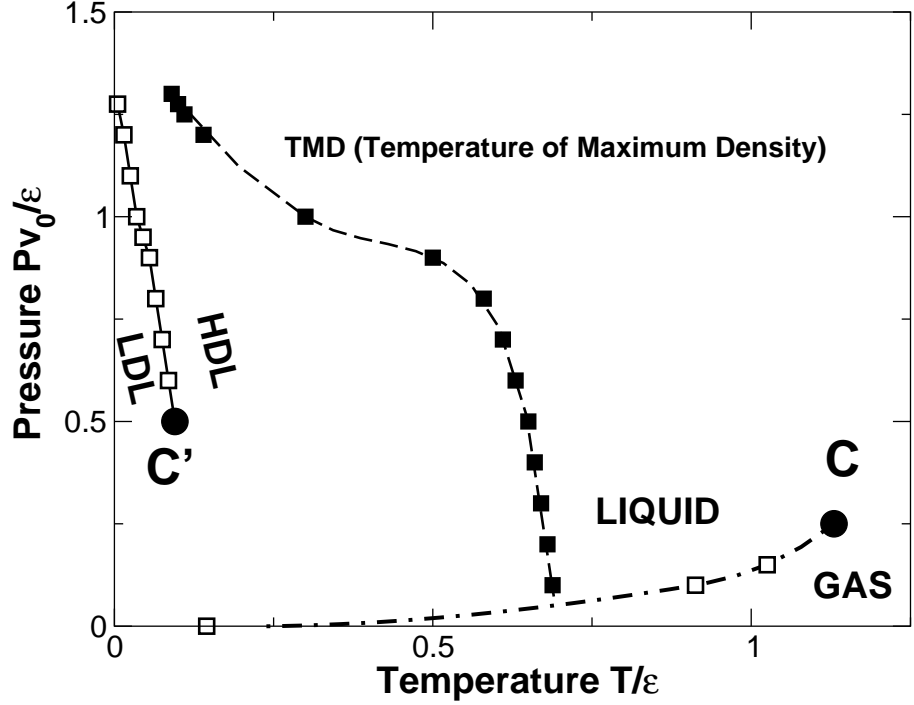}
\caption{The mean field $P$-$T$ phase diagram for the water model with
intra-molecular interaction.
The phase diagram shows the gas-liquid
first-order phase transition (dot-dashed) line ending in the critical point
$C$, the (dashed) line of temperatures of maximum density (TMD)
and the LDL-HDL first-order phase transition (low-$T$ continuous) line
ending in the critical point $C'$ The quantities $v_0$ and $\epsilon$
are internal parameters of the model \cite{fms}. 
\label{fig1}}
\end{figure}

We first check, by theoretical calculations \cite{fs,fsPhysA}
(Figure \ref{fig1}) and computer simulations \cite{fms} (Figure \ref{figMC}),
that our model reproduces the known phase diagram of fluid water, with 
the liquid-gas phase transition ending in the critical point $C$ and
with the TMD line. As shown in the experiments, we 
find that the TMD line decreases with increasing $P$
\cite{ANGELL,Poole97}.  

If we fix the intra-molecular interaction between hydrogen bonds to a positive
values, we find that in the deeply supercooled region the liquid has a liquid-liquid phase
transition ending in the critical point $C'$ and following a line with
negative slope in the $P$--$T$ phase diagram. 
Therefore, we recover the liquid-liquid critical point scenario.

\subsection{Connecting the two interpretations}

At this point it is natural to ask if the liquid-liquid critical point is a 
{\it necessary} consequence of the properties of water. With the
schematic model we use, it is possible to answer this question by
changing the values of (a) the parameters determining the density anomaly
and (b) the parameter determining the correlation among the hydrogen bonds. 

\begin{figure}
\includegraphics[width=12cm]{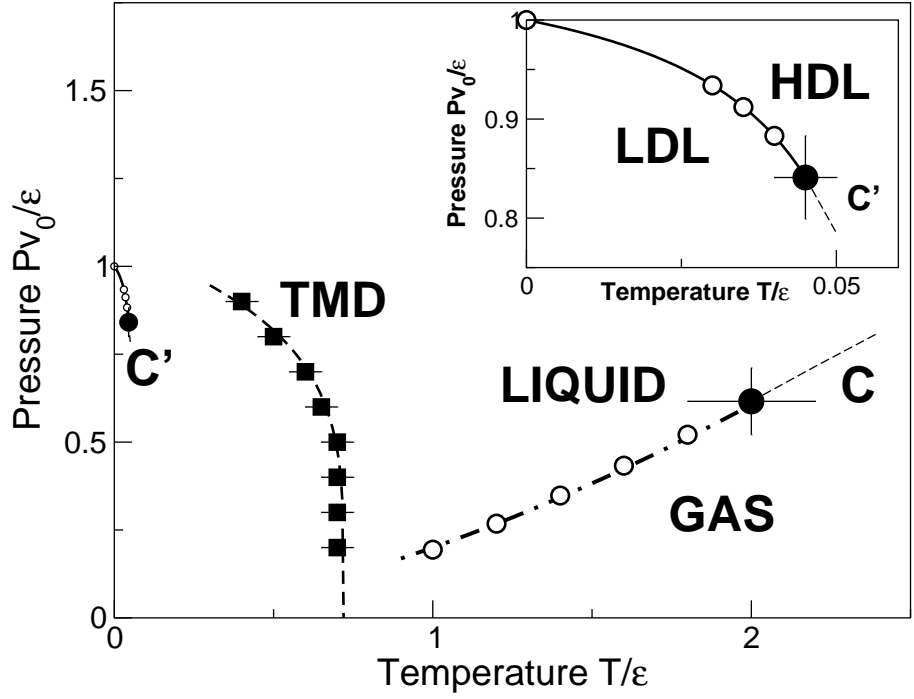}
\caption{
$P$--$T$ phase diagram calculated by computer simulations,
for the model with intra-molecular interaction 
with the symbols as in Fig.\ref{fig1}.
Thin dashed lines indicate the position of maxima of 
$K_T$ (Widom line) emanating from the
critical points $C$ and $C'$. 
Inset: blowup of the HDL-LDL phase transition region.
\label{figMC}}
\end{figure}

We first show that by decreasing the parameter in (b), associated to the
correlation in the hydrogen bond network, the liquid-liquid critical point $C'$ moves to
lower $T$ and higher $P$. When $C'$ ``retraces'' on the line
of the liquid-liquid phase transition, the divergence of the response functions
occurring on this line is replaced by a maximum occurring in a region 
around a locus (the Widom line in Figure \ref{figMC}) in the $P$--$T$
phase diagram.  

The case with the parameter in (b) set to zero, studied in
Ref.~\cite{Sastry96,Rebelo98,LaNave99}, is reached 
with continuity, with $C'$ disappearing at $T=0$ and at the corresponding
pressure on the liquid-liquid phase transition line in Figures \ref{fig1} and \ref{figMC},
leaving behind the Widom line with the maxima of the response functions,
necessary in the singularity-free scenario.

Therefore, we show, first, that the liquid-liquid critical point is, in this model, a
consequence of the hydrogen bond correlation and, second, that  
two scenarios presented above are
related to each other: by decreasing the correlation of the
hydrogen bond network, we pass with continuity from the liquid-liquid phase transition
scenario to the singularity-free interpretation. The latter, within our
schematic model, is only possible in the limiting case of no hydrogen bond
correlation, a situation that is not consistent with the experimental
data for water. 

\subsection{The consequence of the density anomaly}

In all the considered cases, we find that the density anomaly and the TMD line
are not affected by the variation of the parameter in (b), consistent
wit the fact that the these properties are reproduced also when this
parameter is set to zero, i. e. in the singularity-free scenario
\cite{Sastry96,Rebelo98,LaNave99}. 
Instead, it is possible to see that, in our model, the parameter in (a)
determining the density anomaly and the occurrence of the TMD line,
is associated to the local volume expansion caused by the hydrogen bond formation. 
This local expansion is responsible also for the open structure of the
low-density liquid phase. We, therefore, conclude that the liquid-liquid phase
transition is a 
necessary consequence of the density anomaly (due to the open
structure) {\it and} of the hydrogen bond correlation characterizing the
tetrahedral network.

\section{Are these anomalies exclusive properties of water?} 

The change of the local structure of the liquid water, related to its
anomalies,  
has been proposed as a possible mechanism for biological processes
\cite{Authenrieth,WIGGINS90,WIGGINS01,manuel,debenedettibook}.
Therefore, in the debate about the essentiality of water for life
could be relevant to understand which are the liquids that have
density anomaly.

Since we have shown that the density anomaly implies the liquid-liquid phase 
transition when the liquid forms a correlated network of bonds,
it is natural to ask if we can use the
occurrence of a liquid-liquid phase transition as a signal of density anomaly.
For example, recent experiments have shown a liquid-liquid phase transition in
phosphorous \cite{Katayama,Katayama04,Monaco} and in triphenyl phosphite
\cite{Kurita,Kurita04}.

Apart from the clear evidences in phosphorous and triphenyl phosphite,
experiments show data consistent with the
possibility of a liquid-liquid phase transition in
silica \cite{ANGELL,Poole97,Lacks}, carbon \cite{Thiel}, aluminate
liquids \cite{McMillan,Wilding01,Wilding02}, selenium
\cite{Brazhkin98}, and cobalt \cite{Vasin}, among others
\cite{McMillan,Wilding01,Wilding02,McMillan2}.  
Moreover, computer simulations predict a liquid-liquid critical 
point for specific models of
carbon \cite{G}, phosphorous \cite{MORISHITA}, supercooled silica
\cite{ANGELL,Poole97,Saika-Voivod,Sastry-Angell}, and hydrogen
\cite{hydrogen}, besides the commonly used models of water
\cite{Poole,geiger}.

The strategy we follow, to explore if the occurrence of a
liquid-liquid phase transition is {\it sufficient} for the density anomaly in a
liquid, consists in looking at a simple model liquid with the liquid-liquid
critical point and test for the density anomaly. 
To realize this investigation we use a model where the
interaction between molecules depends only on their relative distance
and not on their relative orientation. 

This kind of model have been used 
to represent materials such as liquid metals
\cite{hs1,STILLINGER,Mon,Silbert,Levesque,Kincaid,Cummings,Velasco,Voronel},
colloids or biological solutions \cite{Baksh}. They have been 
proposed also to study anomalous liquids, such as water
\cite{Head-Gordon93,debenedettibook,jagla2,REZA,Sadr99,Scala01,Scala00}.

\begin{figure}
\includegraphics[width=11cm]{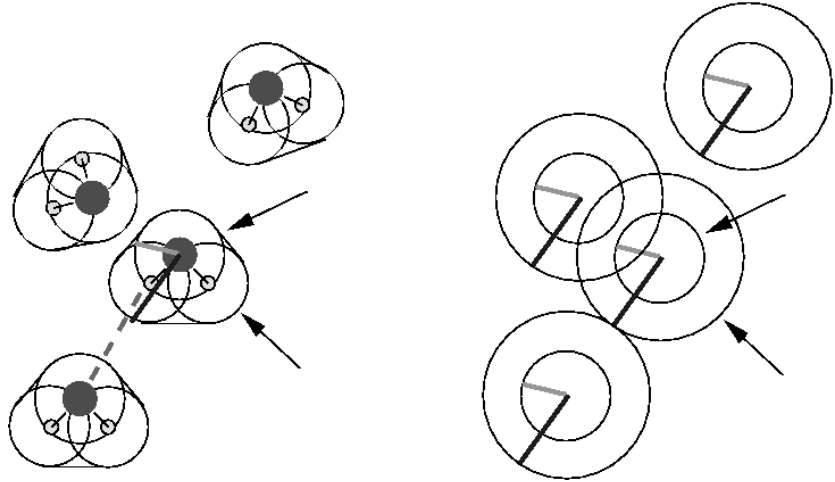}
\caption{
Left: Schematic representation of characteristic distances between water
molecules. The sum of the excluded volumes (open circles) around the
oxygen (large full circle) and the two hydrogens (small gray circles) of H$_2$O
roughly corresponds to the excluded volume of the molecule.
The light-gray short-radius represents the minimum distance between two
molecules without hydrogen bond. The black long-radius represents the distance
between two molecules with hydrogen bond (dashed line). 
Right: The corresponding representation of the molecules as spherical
particles interacting via an isotropic potential. The potential can be
regarded as resulting from an average over the angular part of a more
realistic non-isotropic potential. The isotropic potential has an
impenetrable hard-core (at the short light-gray radius) and a penetrable
soft-core (at the black long-radius). Two spherical particles at the
hard-core distance have overlapping soft-core and
would represent two 
molecules without hydrogen bond and approaching at the minimum distance (upper
arrow). Two spherical particles at the soft-core distance (lower arrow)
would represent two molecules forming a hydrogen bond.
\label{2radiuses}}
\end{figure}

As we have seen, water has an open (at low $P$ and $T$) and a closed
(at high $P$ and $T$) local structure \cite{mishima_2000,Soper-Ricci00}.  
The existence of these two structures with different densities
suggests us a pair interaction with two characteristic distances.  For
example, in water the shortest distance could be associated with the
minimum distance between two non hydrogen bonded molecules (closed
structure). The largest distance would
correspond to the average distance when the hydrogen bond is formed (open structure),
as in Figure \ref{2radiuses}. The tendency
to have an open structure, would correspond to a weak
repulsion at lengths between the shortest and the largest distance.
The energy gain for the hydrogen bond formation would attract the molecules at
the the largest distance. We analyze this simple model with two
characteristic interaction distances by means of numerical
simulations and theoretical calculations.

\subsection{Crystal open structure and polymorphism}

\begin{figure}
\includegraphics[width=12cm]{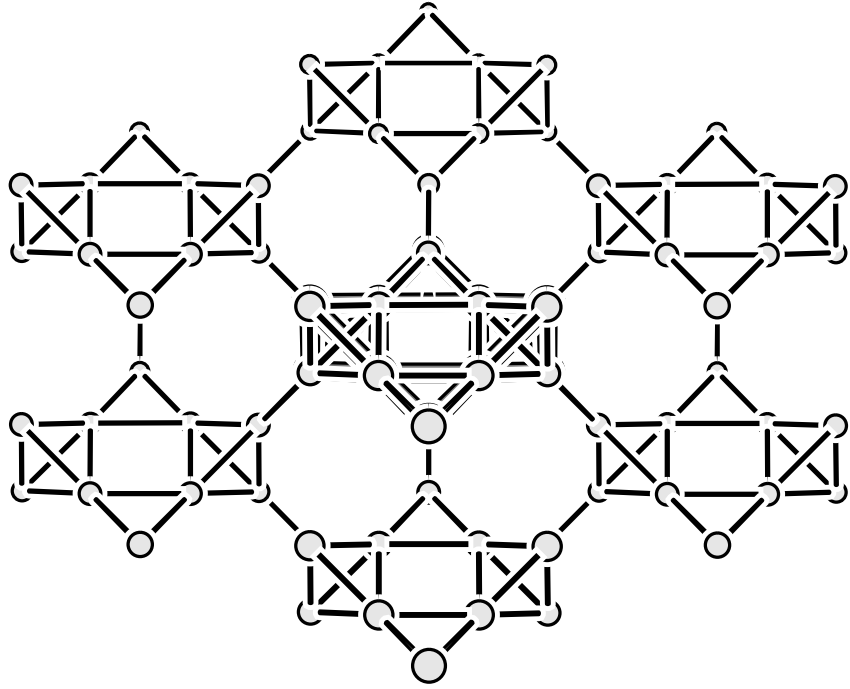}
\caption{
Crystal open structure of the spherical model with two characteristic 
distances.
Bonds connect particles at the attractive distance.
The radius of the
particles is {\it not} in scale with the
distances. Greater particles are closer to the observation point,
reflecting the eight different levels represented in the figure. The
configuration contains 15 cells. The central cell is emphasized by
darker bonds. The cell is formed by 10 particles.
\label{crys}}
\end{figure}

The first surprising result of our simulations is that this simple
model is able to reproduce one of the most relevant features of
water, i.e. a crystal open structure (Figure \ref{crys}) \cite{fmsbs}.
The crystal has a complex unit cell of ten molecules, some at
the shortest distance, other at the largest distance, with 8-fold
and 12-fold symmetries.

This structure appears to be quite stable on a wide range of pressures
and temperatures and is the only one we find by equilibrating the
system with molecular dynamics simulations. 
However, since it does not correspond to the most compact
configuration, we know that by increasing the
pressure the system will ultimately take a (cubic or
hexagonal) close packed form. Therefore, the model 
is able to reproduce the polymorphism typical
of water-like substances. It has, at least, two crystals. A low
density crystal at low and moderate densities and a high-density
crystal at high densities (or pressures). 

\subsection{The liquid-liquid critical point}

\begin{figure}
\includegraphics[width=12cm]{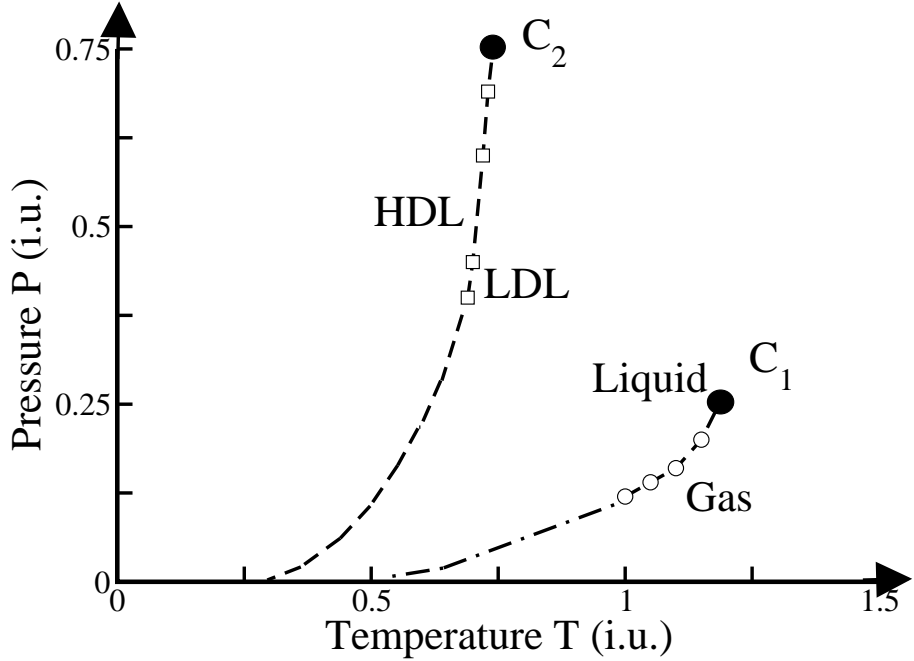}
\caption{
The $P$--$T$ phase diagram for the spherical model with two
characteristic distances, 
showing the gas-liquid critical point $C_1$ and the LDL-HDL critical
point $C_2$. The two critical points (full circles) are located 
at the end of the gas-liquid coexistence (dot-dashed) line 
and the LDL-HDL coexistence (dashed) line, respectively.
Open squares and circles are results of computer simulations.
Lines are schematic guides to the eyes.
Pressure and temperatures are shown in internal units (i.u.), given by
ten times the attractive energy divided by the hard-core volume for
$P$, and the attractive energy divided by the Boltzmann constant for
$T$.  
\label{LL}}
\end{figure}

By supercooling the liquid 
we can avoid the crystal and study the metastable phase diagram below
the melting temperature.
Below the liquid-gas phase transition, ending in the liquid-gas
critical point $C_1$, we observe two liquids in the supercooled phase,
roughly corresponding to 
the two crystals mentioned above (Figure \ref{LL}). 

The two liquids are 
separated by a liquid-liquid phase transition
ending in a critical point $C_2$  \cite{nature,fmsbs,sbfms}.
Therefore, the presence of two length scales is enough to give rise to
a phase diagram with two liquid phases, i.e. two liquids with
different local structure. However, the slope of the
liquid-liquid transition line is positive in the $P$--$T$ phase diagram,
consistent with a HDL with less entropy than the LDL, suggesting that
this model is not describing a water-like system.
Nevertheless, the model is appropriate to test if the liquid-liquid transition is
sufficient for the density anomaly.

\subsection{Condition for the liquid-liquid critical point}

By varying the parameters of the model ruling the repulsion and the
attraction, we find that the liquid-liquid critical point  $C_2$ exists
at positive pressures only in a finite range of parameters because 
its pressure decreases as the its temperature increases. We
rationalize this result by using an approximate theoretical approach 
(a modified van der Waals equation) which qualitatively reproduces the
behavior \cite{sbfms}. Our study shows that not any liquid with two
characteristic interaction distances can have a liquid-liquid phase transition, because
a balance between attraction and repulsion is necessary.

Indeed, when the attraction is too strong the liquid-liquid phase
transition goes to negative pressures, while when the 
attraction is too week, the liquid-liquid phase transition
occurs in the deeply supercooled liquid phase, becoming difficult to
observe, as in the experimental situation of
water or silica~\cite{ms98,fms,Sastry-Angell}.
An extensive investigation, performed with a fast
computational method (the integral equation
approach in the hypernetted-chain approximation) allow us to
quantify better this balance, in terms of the parameters of the
model \cite{fmsbs,mfsbs}.

\subsection{The density behavior}

Once we fix the parameters of the model in such 
a way to satisfy the condition for the liquid-liquid phase transition, we test if
the liquid has an anomalous behavior in density. Surprisingly, we
find, by means of 
thermodynamic integration based on our simulation results, that
the liquid has no density anomaly \cite{fmsbs,sbfms}. 

This result excludes the possibility that the coexistence of two
liquid phases, one with an open local structure and the other with a
closed local structure, in a substance is {\it sufficient} for
determining the anomalous behavior of density and the presence of a
TMD line. Therefore, even if the density anomaly is not an exclusive
property of water, since other substances such as silica show it, the
solely presence of more than a liquid phase does not signal the
occurrence of density maxima.

\section{Outlook} 

Many open questions remain, and many experimental results are
of potential relevance to the task of answering the question about the
importance of water for life. For example, the dynamics of water 
could play a fundamental role in biological processes, such as 
determining the protein folding rate. Hydrophobic collapse and sharp
turns (or bends) in polypeptide chains (groups of 50--100 amino acids)
in the (secondary) structure of proteins
involve the mediation of the water molecules in proximity to the
amino acids \cite{Baron97}.
Protein association could be determined by the the dynamics in the water
hydration layer around appropriate amino acid residue sites \cite{Camacho00}.
Interfacial water could rule the rate of recognition of 
binding sites of proteins by ligands, inhibitors, and
other proteins \cite{Zou02} and the 
water dynamics at the surface of DNA and macromolecules 
is a promising field of research \cite{masaki,Bagchi2005}.

In this context, it is unclear which could be the effect of the 
hypothesized second critical point. Is intruiguig to observe that 
the estimated temperature of 220~K for the liquid-liquid critical
point of water 
\cite{ms98} coincides with the temperature below which 
the enzymatic activity of most proteins ceases
\cite{Rasmussen92,Ferrand93}.
This liquid-liquid phase
transition is probably hindered by inevitable freezing \cite{fms}. Indeed, it
appears that the liquid-liquid phase transition is below, or at least close
to, the glass transition temperature. 
A recent simulation analysis of the orientational dynamics of water at
fixed density \cite{pradeep} has shown that the temperature of
dynamical arrest of the system, 
defined by the (mode coupling) theory \cite{MCT,Gotze92}, is relatively
close in temperature and density to recent estimate of the
liquid-liquid critical 
point \cite{geiger}. 

The anomalous properties of water could be relevant for life in many
ways. For example, trees survive arctic temperatures because
the water in the cell does not freeze, even though the temperature is
below the homogeneous nucleation temperature of -38$^\circ$C. This
effect, due to the confinement of water, is related to the
difficulty for the molecules to form an ordered hydrogen bond tetrahedral
network. 

The correlation of the hydrogen bonds forming the network is, indeed, 
the reason of the occurence of the liquid-liquid phase transition, whose
necessity for water is supported by theoretical and numerical
calculations \cite{fms,slt}.
We find that the position of the liquid-liquid critical point
depends on the strength of the hydrogen bond correlation. 
For example, if we assume
uncorrelated hydrogen bonds \cite{Luzar-Chandler,Luzar96} we find that the liquid-liquid critical
point disappear at $T=0$, giving rise to the 
singularity-free scenario. However, the hypothesis of a vanishing hydrogen bond
correlation does not apply to water \cite{raiteri}, ruling out the 
singularity-free scenario for H$_2$O.

The ability of water to have an open and a closed
structure, determines the existence of polyamorphs, the LDA, the HDA
and the VHDA ice, each in a possible correspondence 
with, respectively, the LDL, the HDL and
the very HDL (VHDL) shown by computer simulations
\cite{b,bs,gss,geiger,w}.
Their different, and anticorrelated, specific
volume and entropy, appears to be the feature that
gives rise to the anomalous properties of water.

However, these properties are not exclusive of water.
Other tetrahedrally-coordinated
liquids have, at low temperature and low pressure, anticorrelated
entropy and specific volume fluctuations. 
Examples of 
such systems are SiO$_2$ and GeO$_2$, known for their
geological and technological importance. Recent simulations show,
for example, that silica and silicon has a liquid-liquid critical point 
\cite{Saika-Voivod, Sastry-Angell}, while experiments show the
occurrence of a liquid-liquid critical point in phosphorous
\cite{Katayama,Katayama04,Monaco} and in triphenyl phosphite
\cite{Kurita,Kurita04}.

Interestingly, some properties of water, such as the 
polymorphism or the existence of a low-density open crystal, 
could be present also in substances without density anomaly, but with two
liquids with different local structures.
Since the change of local liquid structure 
could be relevant in biological processes
\cite{Authenrieth,WIGGINS90,WIGGINS01,manuel,debenedettibook,Bagchi2005}, 
the possibility of a wide class of liquids with this property 
could help in understanding if water is essential for life.

\section*{Acknowledgments}

We thank our collaborators, S. V. Buldyrev, N. Giovambattista,
P. Kumar, G. Malescio, M. I. Marqu\'es, 
F. Sciortino, A. Skibinsky, and M. Yamada.
We thank Chemistry Program CHE 0096892 and CHE0404673 for support. 
G. F. thanks the
Spanish Ministerio de Educaci\'on y Ciencia (Programa Ram\'on y Cajal and
Grant No. FIS2004-03454).

\newpage 

\end{document}